\documentclass[11pt,a4paper]{article}

\usepackage[utf8]{inputenc}
\usepackage[T1]{fontenc}
\usepackage{lmodern}
\usepackage[english]{babel}
\usepackage{microtype}
\usepackage{amsmath,amssymb,amsthm}
\usepackage{mathtools}
\usepackage{graphicx}
\usepackage{booktabs}
\usepackage{tabularx}
\usepackage{multirow}
\usepackage{array}
\usepackage{caption}
\usepackage[margin=2.5cm]{geometry}
\usepackage[hidelinks]{hyperref}
\hypersetup{
  pdftitle={Information-Theoretic Reliability is Robust to Analytic Choice: A 24-Specification Multiverse on Public Cognitive Test-Retest Data},
  pdfauthor={Maria Westrin},
  pdfkeywords={reliability paradox, mutual information, ICC, multiverse analysis, test-retest, cognitive tasks}
}
\usepackage{xcolor}
\usepackage{enumitem}
\usepackage{natbib}
\usepackage{orcidlink}
\usepackage{etoolbox}
\usepackage{titling}
\usepackage{abstract}
\usepackage{authblk}
\title{\bfseries Information-Theoretic Reliability is Robust to Analytic Choice: \\
A 24-Specification Multiverse on Public Cognitive Test--Retest Data}

\author[1]{Maria Westrin\thanks{Corresponding author: \texttt{maria.westrin0293@stud.hkr.se}; ORCID: \href{https://orcid.org/0009-0005-8730-0931}{0009-0005-8730-0931}}\,\orcidlink{0009-0005-8730-0931}}
\affil[1]{Kristianstad University, Kristianstad, Sweden}
\date{May 17, 2026}

\newcommand{\NLRdelta}{\mathrm{NLR}_{\Delta}}
\newcommand{\NLRratio}{\mathrm{NLR}_{\rho}}
\newcommand{\MI}{\mathrm{MI}}
\newcommand{\ICC}{\mathrm{ICC}}
\newcommand{\BCa}{\mathrm{BC_a}}

\newcommand{\E}{\mathbb{E}}

\begin{document}

\maketitle

\thispagestyle{empty}

\begin{center}
\small
\textit{Preprint. The article text and figures are licensed under
\href{https://creativecommons.org/licenses/by-nc-nd/4.0/}{CC BY-NC-ND 4.0}
(no derivatives permitted).
The accompanying software and methodology (MixMind Reliability Framework v2.0.0,
Zenodo DOI \texttt{10.5281/zenodo.20207371}) are governed by the
\textbf{MixMind Research Use and No-Derivatives License v1.0}, which prohibits
commercial use, derivative works, and substitute implementations without
written permission. See the repository \texttt{LICENSE} file for full terms.}
\end{center}

\vspace{1em}

\begin{abstract}
\noindent
\textbf{Background.}
The \emph{reliability paradox} \citep{hedge2018reliability} describes the empirical
observation that cognitive tasks producing robust group-level effects often
yield poor between-individual reliability. Existing approaches rely
predominantly on the intraclass correlation coefficient (ICC), which
captures only linear, second-moment dependence between test and retest.
\textbf{Methods.}
We introduce a normalized, information-theoretic complement to ICC,
$\NLRdelta$, defined as the difference between empirically estimated
mutual information and the analytic Gaussian baseline implied by the
test--retest correlation. We pair $\NLRdelta$ with $\ICC(2,1)$,
bias-corrected and accelerated ($\BCa$) bootstrap intervals
(\(B = 5000\); no percentile fallback was triggered in the present analysis), Benjamini--Hochberg false discovery rate (FDR) control,
and a $24$-cell multiverse over the KSG nearest-neighbour parameter,
correlation method, and minimum-sample threshold. The full pipeline is
governed by pre-specified claim contracts, content-addressed
provenance, and SHA-256-verified raw data ingestion, and is released
as the MixMind Reliability Framework (Software version 2.0.0,
Result version v1.2.2; \citealp{westrin2026mixmind}). We re-analyse the
public test--retest data of \citet{hedge2018reliability} and
\citet{clark2017cognitive}.
\textbf{Results.}
Across $50$ estimable primary measures from the Flanker, Stroop,
Stop-Signal, Go/No-Go, and Posner task families (Visual Working
Memory was pre-specified but non-estimable under the default
specification), the median
$\NLRdelta$ is $-0.138$ nats (interquartile range
$[-0.257,\, -0.034]$); the median two-sided $\BCa$ confidence interval
width is $0.558$ nats, implying a median minimum detectable effect of
$\approx 0.279$ nats. Zero of $50$ primary measures exceed the headline
rule (one-sided $\BCa$ lower bound $> 0$); the smallest
Benjamini--Hochberg-adjusted $q$-value is $0.995$. The companion
$\ICC(2,1)$ analysis recovers the classical reliability paradox pattern
(median $\ICC = 0.615$, range $[0.022,\, 0.764]$). The $24$-specification
multiverse yields $0$ of $1{,}200$ estimable cells passing the headline
rule (median $\NLRdelta = -0.134$).
\textbf{Conclusions.}
On these two public datasets, replacing or augmenting ICC with an
information-theoretic reliability measure does not rescue cognitive
tasks from the reliability paradox. The robust null is invariant to the
analytic choices examined here. We release the full pipeline, raw-data
hashes, and contracts to enable exact replication and to invite
extension to other datasets and tasks.

\end{abstract}

\section{Introduction}
\label{sec:intro}

\subsection{The reliability paradox}

A well-replicated tension in cognitive science is that experimental
tasks designed to produce large, robust group-level effects often fail
to produce reliable individual-difference scores
\citep{hedge2018reliability,parsons2019psychological,enkavi2019large,zorowitz2023improving}.
\citet{hedge2018reliability} reported that classical paradigms
(Flanker, Stroop, Stop-Signal, Go/No-Go, Posner cueing, Visual Working
Memory) yielded test--retest intraclass correlations
($\ICC$s) frequently below conventional thresholds, despite producing
group-level Cohen's $d$ effect sizes well above $1.0$. They termed this
gap the \emph{reliability paradox}: paradigms optimised for
within-subject contrasts minimise between-subject variance, which is
exactly the variance individual-differences research requires
\citep{hedge2018reliability}.

The methodological response has emphasised two routes. First,
\emph{measurement-model upgrades} replace single trial-mean scores
with hierarchical Bayesian or generative models, recovering reliability
by jointly modelling within-subject noise and between-subject signal
\citep{haines2023learning,clayson2024psychometric}. Second,
\emph{task redesign} calibrates trial composition (e.g., difficulty
mixing) so that between-subject variance is restored without sacrificing
group-level effect detection \citep{kucina2023calibration}. Recent work
clarifies that the paradox extends beyond reliability per se: low
reliability also attenuates group-level effect sizes
\citep{karvelis2025clarifying}.

\subsection{The unmet need: a non-linear reliability complement}

Existing reliability estimators almost universally rely on linear,
second-moment relationships between test and retest. The intraclass
correlation coefficient $\ICC(2,1)$ \citep{shrout1979intraclass,
mcgraw1996forming, koo2016guideline} captures the proportion of total
variance attributable to consistent between-subject differences,
treating the test--retest joint distribution as approximately
bivariate Gaussian. If a task's true reliability is non-linear, or if
its score distribution has heavy tails, $\ICC$ may be \emph{both}
biased downward and unable to express the structure that does remain.

Information theory offers a strictly more general dependence measure.
The \emph{mutual information} between test ($X_1$) and retest
($X_2$),
\begin{equation}
\MI(X_1; X_2) \;=\; H(X_1) + H(X_2) - H(X_1, X_2),
\label{eq:mi-defn}
\end{equation}
is non-negative, equals zero if and only if $X_1 \perp X_2$, and
captures dependence of all orders \citep{shannon1948mathematical,
cover2006elements}. Mutual information is invariant under monotone
transformations of either variable; under bivariate normality with
Pearson correlation $\rho$, it admits the closed form
\citep{cover2006elements}
\begin{equation}
\MI_{\mathrm{Gauss}}(\rho) \;=\; -\tfrac{1}{2}\,\log\!\bigl(1 - \rho^2\bigr).
\label{eq:mi-gauss}
\end{equation}

A reliability index that exploits this generality should: (i)
estimate $\MI(X_1; X_2)$ non-parametrically; (ii) calibrate that
estimate against the Gaussian baseline implied by the observed $\rho$,
so the estimator distinguishes genuine non-linear structure from
ordinary linear correlation; (iii) admit confidence intervals robust
to small samples and skew; and (iv) report the analyst's pipeline of
choices transparently. Within psychology, mutual-information-based
reliability has been developed only for latent-class diagnostic
models \citep{wang2018mutual}, not for response-time-based cognitive
paradigms. The present paper fills this gap.

\subsection{Contributions}

\begin{enumerate}[leftmargin=*]
  \item We define $\NLRdelta$, a normalized, information-theoretic
        reliability index for test--retest data on continuous outcomes.
        $\NLRdelta$ is the difference between an empirical mutual
        information estimate (Kraskov--St{\"o}gbauer--Grassberger;
        \citealp{kraskov2004estimating}) and the analytic Gaussian
        baseline of equation~\eqref{eq:mi-gauss}, expressed in nats.
        Positive values indicate non-linear structure beyond what the
        observed correlation explains; values $\leq 0$ indicate that
        the bivariate Gaussian baseline already captures the available
        dependence.
  \item We pair $\NLRdelta$ with $\ICC(2,1)$, bias-corrected and
        accelerated ($\BCa$) bootstrap confidence intervals
        \citep{efron1987better, diciccio1996bootstrap},
        Benjamini--Hochberg false discovery rate control
        \citep{benjamini1995controlling}, and a multiverse analysis
        \citep{steegen2016increasing} over plausible analytic
        choices.
  \item We embed the full estimation pipeline in
        \emph{contracts}: machine-readable JSON specifications that
        bind, before analysis, the set of measures admissible for
        headline claims (the \emph{primary tier}), the set admissible
        only for sensitivity analyses, the headline test rule, and the
        names and definitions of all reliability quantities computed.
        Each release records SHA-256 hashes of input archives and
        outputs, the run mode, the bootstrap budget, and the full
        Python dependency lock.
  \item We apply the framework to two public test--retest datasets:
        the cognitive-task battery of \citet{hedge2018reliability}
        (Flanker, Stroop, Stop-Signal, Go/No-Go, Posner cueing, and
        Visual Working Memory) and the working-memory training data
        of \citet{clark2017cognitive}. Across $50$ estimable primary
        measures and a $24$-cell multiverse, we observe a robust null
        on the headline rule.
  \item All code, contracts, raw-data hashes, multiverse outputs, and
        a verifier script are released. The release passes a $16$-check
        promotion gate (R1--R16), enabling third-party reproduction.
\end{enumerate}

\subsection{Scope and what we do not claim}

This paper is an applied-statistics contribution, not an
estimator-theoretic one. We do not derive new properties of mutual
information or of $\ICC$. We do not claim that no cognitive task can be
made reliable; the literature on hierarchical generative models and
task calibration shows that some can
\citep{haines2023learning,kucina2023calibration}. We do not claim that
$\NLRdelta$ \emph{should} replace $\ICC$; we report it alongside
$\ICC(2,1)$ precisely to enable readers to draw their own
methodological conclusions. The contribution is the combination, the
multiverse, the contracts-based reproducibility framework, and the
empirical evidence that, on these two public datasets, an
information-theoretic complement does not lift the paradox.

\section{Methods}
\label{sec:methods}

\subsection{Data}
\label{sec:data}

We re-analysed two publicly available test--retest datasets. Raw archives
were downloaded once, verified by SHA-256 hash against the values recorded
in the release provenance, and ingested through a single auditable
extraction routine.

\paragraph{\citet{hedge2018reliability}.}
Two sessions per participant on six classical cognitive tasks
(Flanker, Stroop, Stop-Signal, Go/No-Go, Posner cueing, Visual Working
Memory). After exclusions for incomplete sessions and non-finite
trial-level data, the canonical long-format dataset comprised
$15{,}034$ rows. Trials below $200$\,ms or above $5000$\,ms were
excluded prior to within-subject aggregation, following the original
authors' processing.

\paragraph{\citet{clark2017cognitive}.}
A working-memory training trial in healthy older adults; we used only
the pre/post test--retest measures relevant to the present analysis.
The contract assigns the Clark dataset primarily to the
\emph{sensitivity tier} (see Section~\ref{sec:contracts}); the locked
registry also includes a single Clark-derived working-memory aggregate
(\texttt{clark\_vwm\_kavg}) in the \emph{primary tier}, which was
ingested and hash-verified but proved non-estimable under the default
specification ($n < n_{\min}$). No finite Clark estimate therefore
enters the headline analysis; the remaining 24 Clark endpoints are
reported as sensitivity-tier transparency only.

\paragraph{Provenance.}
The release records SHA-256 hashes of both raw archives,
$902\mathrm{a}00\ldots$ for the Hedge archive and
$928742\mathrm{b}1\ldots$ for the Clark archive. Re-running the
pipeline on a different copy of either archive produces a hash
mismatch and aborts.

\subsection{Measure tiers (claim set)}
\label{sec:contracts}

We pre-specified a measure registry partitioning every reliability
estimate into one of four tiers (Table~\ref{tab:tiers}):

\begin{table}[t]
\centering
\caption{Pre-registered measure tiers. Counts are immutable across the
release; only \emph{primary} measures support headline claims.}
\label{tab:tiers}
\small
\begin{tabularx}{\linewidth}{lrX}
\toprule
\textbf{Tier} & \textbf{$N$} & \textbf{Role} \\
\midrule
canonical    & 49 & Long-form measure registry; defines the analytic
                    universe. \\
primary      & 56 & Eligible for headline reliability claims. \\
sensitivity  & 24 & Reported for transparency; not eligible for
                    headline claims. \\
descriptive  &  1 & Reported for context; no inferential role. \\
excluded     &  1 & Defined but not used. \\
\midrule
full pipeline & 82 & Total measures touched by the pipeline,
                     including derivatives. \\
\bottomrule
\end{tabularx}
\end{table}

The \emph{primary} tier defines the inferential universe over which
headline claims are tested. The contract is enforced at run time: any
attempt to derive a headline claim from a non-primary measure aborts
the pipeline.

\subsection{Reliability estimators}

\subsubsection{Intraclass correlation, \texorpdfstring{$\ICC(2,1)$}{ICC(2,1)}}

Following \citet{shrout1979intraclass} and \citet{mcgraw1996forming},
we estimate $\ICC(2,1)$, the single-rater absolute-agreement
two-way random-effects intraclass correlation, on the within-subject
test--retest score pairs for each measure. Sessions are modelled as
exchangeable measurement occasions (random effects) under the
absolute-agreement variance decomposition. We report $\ICC$ point
estimates with $F$-distribution-based confidence intervals;
$\ICC(3,1)$ (mixed-effects, fixed-rater) is reported in the
per-measure output as a sensitivity check only.

\subsubsection{Mutual information, KSG estimator}

We estimate $\MI(X_1;X_2)$ non-parametrically using the
$k$-nearest-neighbour estimator of
\citet[Eq.~(8)]{kraskov2004estimating} (KSG, variant 1):
\begin{equation}
\widehat{\MI}_{\mathrm{KSG}}(X_1;X_2) \;=\;
  \psi(k) - \E\!\left[\,\psi(n_{x_1}+1) + \psi(n_{x_2}+1)\,\right]
  + \psi(N),
\label{eq:ksg}
\end{equation}
where $\psi$ is the digamma function, $N$ is the number of paired
observations, and $n_{x_1}, n_{x_2}$ count points within the
$\ell^\infty$ neighbourhood defined by the $k$-th nearest joint
neighbour. The KSG estimator is asymptotically unbiased and has been
extensively validated for low-dimensional joint distributions
\citep{ross2014mutual}. We use $k=4$ as the primary specification and
$k\in\{3,4,5,6\}$ in the multiverse.

\subsubsection{Gaussian baseline}
\label{sec:methods-baseline}

For each measure we compute the test--retest Pearson correlation
$\hat\rho$ and the corresponding analytic Gaussian mutual information
of equation~\eqref{eq:mi-gauss}:
\begin{equation}
\widehat{\MI}_{\mathrm{Gauss}} \;=\; -\tfrac{1}{2}\,\log\!\bigl(1 -
\hat\rho^{\,2}\bigr).
\end{equation}
This is the mutual information that the joint sample \emph{would} have
if it were exactly bivariate Gaussian with the observed correlation.
\textbf{In the multiverse, when
\texttt{corr\_method = spearman} (Section~\ref{sec:multiverse-analysis}), we
substitute Spearman's $\rho_s$ in place of Pearson's $\rho$ in the
formula above. This is properly described as a rank-correlation
sensitivity baseline rather than strict bivariate Gaussian MI; it
tests robustness of the headline finding to monotonic but non-linear
rank structure, but does not correspond to a closed-form MI of any
specific distribution.}

\subsubsection{\texorpdfstring{$\NLRdelta$ and $\NLRratio$}{NLR-Delta and NLR-rho}}

We define the headline reliability index
\begin{equation}
\boxed{\;
\NLRdelta \;\equiv\; \widehat{\MI}_{\mathrm{KSG}} -
                   \widehat{\MI}_{\mathrm{Gauss}}
\;}\quad\text{(nats)}
\label{eq:nlr-delta}
\end{equation}
and a sensitivity-only ratio,
\begin{equation}
\NLRratio \;\equiv\; \widehat{\MI}_{\mathrm{KSG}} \,/\,
                     \widehat{\MI}_{\mathrm{Gauss}}.
\label{eq:nlr-ratio}
\end{equation}
$\NLRdelta$ is on a natural scale (nats), additive across independent
information sources, and $\leq 0$ when no non-linear structure beyond
the linear correlation is detectable. $\NLRratio$ is reported
descriptively only. The contract pre-registers $\NLRdelta$ as the sole
headline quantity; $\NLRratio$ is excluded from inferential use.

\subsection{Confidence intervals: bias-corrected and accelerated bootstrap}

We construct $\BCa$ confidence intervals for $\NLRdelta$ following
\citet{efron1987better} and \citet{diciccio1996bootstrap}. Subjects
are resampled with replacement; for each replicate we recompute the
KSG estimate, the Gaussian baseline, and their difference. The
$\BCa$ correction adjusts both for median bias and for skewness of
the bootstrap distribution; for cases in which the acceleration
estimate is undefined, the implementation falls back to percentile
intervals (\texttt{brain/bootstrap.py}, lines 273--278). For the
present analysis, no fallback was triggered: all 50 estimable primary
measures used standard $\BCa$ intervals. The default replicate
count is $B = 5000$.

\subsection{Headline rule}

The contract pre-registers a single headline test:
\begin{quote}
\itshape
A measure passes the headline rule if and only if the lower bound of
its two-sided $95\%$ $\BCa$ confidence interval on $\NLRdelta$ is
strictly greater than zero.
\end{quote}
\noindent
This is a one-sided test against $H_0\!: \NLRdelta \leq 0$ at the
$2.5\%$ level. We additionally report $p$-values from the bootstrap
distribution and Benjamini--Hochberg-adjusted $q$-values
\citep{benjamini1995controlling} across the primary tier; FDR control
is applied at $q^* = 0.05$.

\subsection{Multiverse analysis}
\label{sec:multiverse-analysis}

To assess robustness to defensible analytic choices we pre-specified
a $24$-cell multiverse over three axes:
\begin{itemize}[leftmargin=*]
\item $k$ (KSG nearest-neighbour parameter): $\{3,4,5,6\}$;
\item correlation method for the Gaussian baseline:
      $\{\text{Pearson},\ \text{Spearman}\}$;
\item minimum sample size per measure $n_{\min}$:
      $\{10, 15, 20\}$.
\end{itemize}
The full grid is the Cartesian product
($4\times 2\times 3 = 24$ specifications). For each specification we
recompute every primary measure, generate $5000$ $\BCa$ bootstrap
replicates, and apply the headline rule. The multiverse passes if at
least one (specification, measure) cell exceeds the rule.

\subsection{Reproducibility}
\label{sec:repro}

The release ships with a $16$-check promotion gate (R1--R16) covering
directory structure, contract integrity, claim-count immutability,
output schema validation, raw-data hash verification, dependency lock
correctness, ingest provenance, and absence of synthetic data.
Final-mode promotion requires all $16$ checks to pass. Provenance JSON
captures Python ($3.12.10$), NumPy ($2.4.4$), SciPy ($1.15.2$), pandas
($2.2.3$), the bootstrap budget ($B = 5000$), the run mode (final),
and the SHA-256 hashes of all primary outputs. The Linux container
recipe and a Windows note on UTF-8 console encoding are included.

\section{Results}
\label{sec:results}

\subsection{Primary tier}

Of the $56$ pre-specified primary measures, $50$ were estimable;
six were retired with insufficient sample sizes
(\texttt{insufficient\_n}). All $50$ estimable measures came from
\citet{hedge2018reliability} (Table~\ref{tab:tasks}); the single
\texttt{primary} measure from \citet{clark2017cognitive} fell below
$n_{\min}$ at the default specification.

\begin{table}[t]
\centering
\caption{Measure counts by task family from \citet{hedge2018reliability}.
``Estimable'' counts the measures that met the default-specification
minimum sample size ($n_{\min} = 10$) and were therefore included in
the headline analysis. Six of the 56 pre-specified primary measures
were retired with insufficient paired observations.}
\label{tab:tasks}
\small
\begin{tabular}{lrrr}
\toprule
\textbf{Task family} & \textbf{Pre-registered} & \textbf{Estimable} & \textbf{Median $n$} \\
\midrule
Flanker        & 16 & 16 & 53.5 \\
Stroop         & 16 & 16 & 53.5 \\
Stop-Signal    & 12 & 12 & 53.5 \\
Go/No-Go       &  6 &  4 & 53.5 \\
Posner         &  5 &  2 & 40 \\
Visual WM      &  1 &  0 & --- \\
\midrule
Total          & 56 & 50 & --- \\
\bottomrule
\end{tabular}
\end{table}

\paragraph{$\NLRdelta$ summary.}
The $50$-measure distribution of $\NLRdelta$ is centred below zero:
median $-0.138$ nats, interquartile range $[-0.257,\,-0.034]$, range
$[-1.07,\,0.098]$. The median two-sided $\BCa$ confidence interval
width is $0.558$ nats, implying a median minimum detectable effect
size (half-width) of $\approx 0.279$ nats. \emph{Zero} of $50$
measures pass the pre-specified headline rule. The smallest
Benjamini--Hochberg $q$-value is $0.995$;
no measure approaches conventional FDR thresholds.

\paragraph{$\ICC(2,1)$ summary.}
The same $50$ measures yield an $\ICC(2,1)$ distribution with median
$0.615$, interquartile range $[0.456,\,0.709]$, and range
$[0.022,\,0.764]$. This pattern reproduces the classical
\citet{hedge2018reliability} result: tasks engineered for large
within-subject contrasts show modest, frequently sub-threshold
between-subject reliability.

\subsection{Multiverse}

The $24$-specification multiverse yielded $1{,}344$ (specification,
measure) cells; $1{,}200$ were estimable and $144$ fell below $n_{\min}$
in their specification. \emph{Zero of $1{,}200$} estimable cells pass
the headline rule. The multiverse-wide median $\NLRdelta$ is $-0.134$
nats (interquartile range $[-0.222,\,-0.042]$).
Table~\ref{tab:multiverse} summarises the marginal distributions.

\begin{table}[t]
\centering
\caption{Multiverse marginals: median $\NLRdelta$ (nats) and pass count
across the $24$ specifications, by axis. \emph{None} of the $1{,}200$
estimable cells passed the headline rule. Values are reported to three
decimal places, taken from the authoritative multiverse summary
(\texttt{reports/AUTHORITATIVE\_multiverse\_summary\_2026-05-12.json}).
All three $n_{\min}$ levels yield identical marginal medians because
the estimable subset of measures is invariant across these thresholds
for the Hedge dataset.
The Pearson rows use the closed-form bivariate Gaussian baseline;
the Spearman rows use a rank-correlation sensitivity baseline (see
Section~\ref{sec:methods-baseline}).}
\label{tab:multiverse}
\small
\begin{tabular}{llrr}
\toprule
\textbf{Axis} & \textbf{Level} & \textbf{Median $\NLRdelta$} & \textbf{Pass / OK} \\
\midrule
$k$ (KSG)        & $3$       & $-0.147$ & 0 / 300 \\
                 & $4$       & $-0.128$ & 0 / 300 \\
                 & $5$       & $-0.134$ & 0 / 300 \\
                 & $6$       & $-0.118$ & 0 / 300 \\
\midrule
Correlation      & Pearson   & $-0.136$ & 0 / 600 \\
                 & Spearman  & $-0.131$ & 0 / 600 \\
\midrule
$n_{\min}$       & $10$      & $-0.134$ & 0 / 400 \\
                 & $15$      & $-0.134$ & 0 / 400 \\
                 & $20$      & $-0.134$ & 0 / 400 \\
\bottomrule
\end{tabular}
\end{table}

\subsection{Distributional view}

\begin{figure}[t]
\centering
\includegraphics[width=0.95\linewidth]{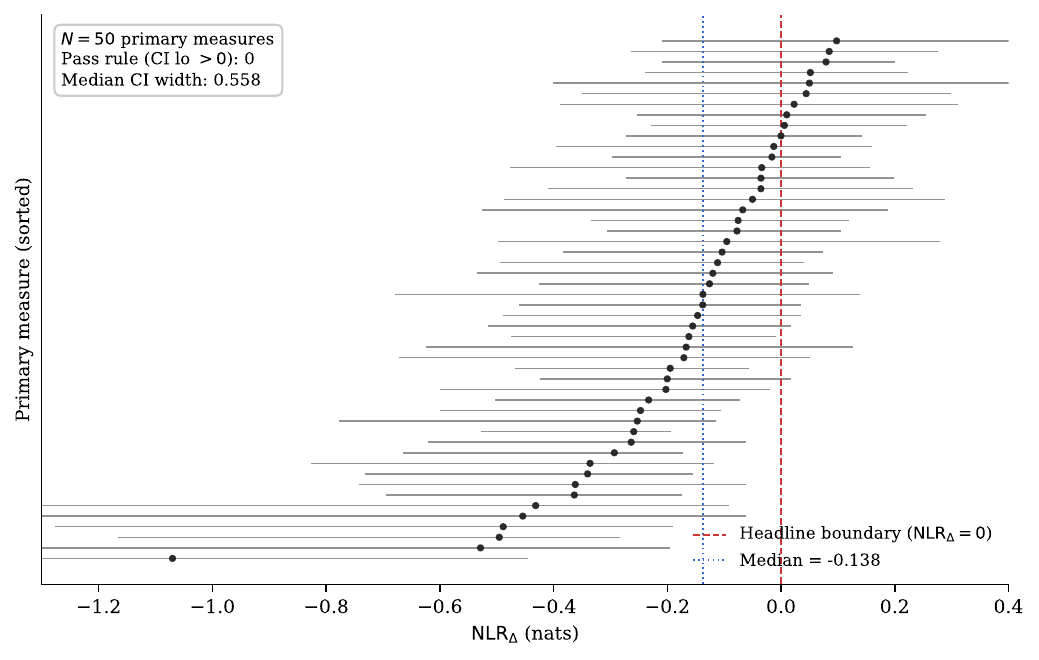}
\caption{Per-measure $\NLRdelta$ (nats) with two-sided $95\%$ $\BCa$
confidence intervals, sorted by point estimate. The dashed vertical
line is the headline boundary ($\NLRdelta = 0$). Zero of $50$
intervals lie strictly to the right of the boundary.}
\label{fig:nlr_dist}
\end{figure}

Figure~\ref{fig:nlr_dist} shows the per-measure $\NLRdelta$ point
estimates with $\BCa$ confidence intervals, ordered. Every interval
crosses or lies below zero; the density mass sits clearly to the left
of the headline boundary.

\begin{figure}[t]
\centering
\includegraphics[width=0.95\linewidth]{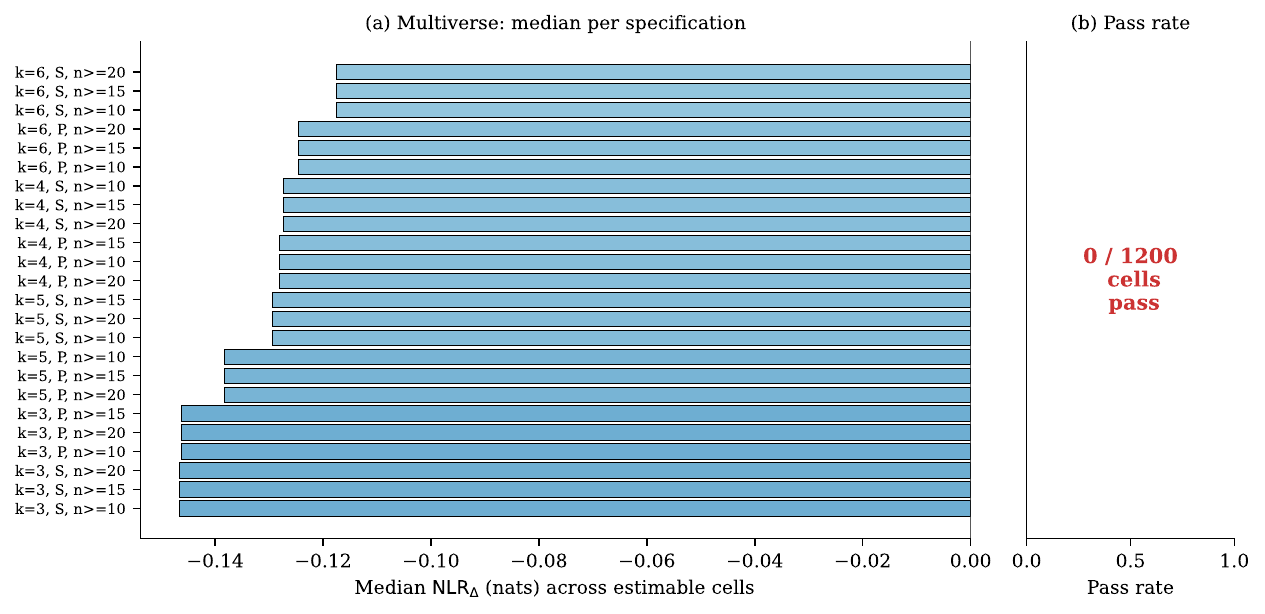}
\caption{Multiverse summary. For each of the 24 specifications
($k \in \{3,4,5,6\}$, $\text{corr} \in \{\text{Pearson},
\text{Spearman}\}$, $n_{\min} \in \{10, 15, 20\}$), the figure
displays the proportion of estimable primary measures that pass the
headline rule ($\BCa$ CI lower bound $> 0$). The pass proportion is
uniformly zero across all 24 specifications.}
\label{fig:multiverse}
\end{figure}

Figure~\ref{fig:multiverse} shows the multiverse view: the proportion
of estimable cells passing the headline rule across the
$24$ specifications. The proportion is uniformly zero.

\section{Discussion}
\label{sec:discussion}

\subsection{What the null means}

On these two public datasets, replacing or augmenting $\ICC(2,1)$ with
a normalized, information-theoretic complement does not rescue
cognitive tasks from the reliability paradox. The result is robust
across $24$ defensible analytic specifications, two correlation
metrics, and three minimum-sample thresholds. We find no evidence for positive excess nonlinear test--retest
dependence beyond the Gaussian correlation baseline within the
detectable range (MDE $\approx 0.279$ nats at the available sample
sizes). The observed negative population median ($-0.138$ nats) is
consistent with the documented KSG-1 finite-sample bias
\citep{gao2015estimating} and should not be interpreted as evidence
of sub-Gaussian dependence. We cannot exclude effects below the
detection threshold; whatever individual-difference signal is present
at the detectable scale is already expressed by $\rho$ and therefore
by $\ICC$.

This is consistent with, and complements, recent work showing that the
paradox stems from the design properties of the tasks
\citep{kucina2023calibration} and from measurement-model choices
\citep{haines2023learning} rather than from a missing higher-order
moment in the marginal score distribution.

\subsection{What we are not saying}

\begin{itemize}[leftmargin=*]
\item We are not saying $\NLRdelta$ is a poor reliability index.
      A null on \emph{these} datasets and \emph{these} specifications
      is informative; it is not generalisable to other tasks, modalities,
      or measurement models.
\item We are not saying $\ICC(2,1)$ is sufficient. The classical
      reliability-paradox finding stands: $\ICC$ values for these
      tasks are, on the median, modest.
\item We are not advocating against hierarchical Bayesian or generative
      modelling. The literature is clear that those approaches restore
      reliability where the present approach does not
      \citep{haines2023learning,clayson2024psychometric}.
\end{itemize}

\subsection{Limitations}

\paragraph{Ties and bounded supports.}
KSG-1 is a continuous nearest-neighbour mutual information estimator
and is not formally calibrated for measures with substantial tied
values or restricted supports. The accuracy-based endpoints in the
present analysis (e.g., proportion-correct on the Flanker, Stroop,
and Go/No-Go congruency conditions) can exhibit ties when small
samples ($n \approx 47$--$60$) generate few unique values on the
$[0, 1]$ accuracy scale. We do not apply jitter or discrete--continuous
mixture estimators; this is a known limitation (see
\texttt{docs/KNOWN\_LIMITATIONS.md} L1 in the released repository).
Reaction-time-based endpoints are not affected by this concern.

The principal limitation is the dataset scope: two public archives,
one of which (Clark) supplies only sensitivity-tier measures under
the present contract. The KSG estimator with $k\in\{3,4,5,6\}$ is
appropriate for the sample sizes here ($n \approx 60$ per measure)
but cannot recover non-linear structure that requires substantially
larger $N$ to detect at the bivariate level. The contract excludes
trial-level approaches; a hierarchical extension that estimates
mutual information directly from trial-by-trial response
distributions is an important open direction.

\subsection{Open invitation}

The full pipeline, contracts, raw-data hashes, and multiverse outputs
are released under the MixMind Research Use and No-Derivatives
License v1.0 (see \citealp{westrin2026mixmind} and the repository
\texttt{LICENSE} file). We invite extensions to other
test--retest archives (e.g., the Many Labs cognitive-task replication
data; \citealp{enkavi2019large}) and to other reliability indices.
Replicating the present pipeline against a third dataset that
\emph{does} contain meaningful non-linear test--retest structure would
both validate $\NLRdelta$ as a positive instrument and enrich the
empirical map of when, if ever, the reliability paradox lifts.

\section*{Open Science Statement}

\paragraph{Data availability.}
Both datasets are publicly archived by their original authors:
\citet{hedge2018reliability} via the original article's supplementary
materials; \citet{clark2017cognitive} via the PLOS ONE article record.
SHA-256 hashes of the exact archives used are recorded in the release
provenance.

\paragraph{Code availability.}
The complete analysis pipeline (the MixMind Reliability Framework,
Software version 2.0.0, Result version v1.2.2;
\citealp{westrin2026mixmind}) is permanently archived on Zenodo with
DOI \texttt{10.5281/zenodo.20207371} (\url{https://doi.org/10.5281/zenodo.20207371}).
Selected supporting files (license, citation, and reproducibility
guide) are included as ancillary files (\texttt{anc/}) with this
preprint; the full repository, containing all source code,
pre-specified claim contracts, the verifier script, the multiverse
engine, raw-data SHA-256 fingerprints, and full execution
provenance, is on Zenodo at the DOI above. The release passes 16/16
automated verification checks and 9/9 release-gate checks. A
byte-identical reproduction of the authoritative
\texttt{per\_measure\_results.csv} (after newline normalisation) and
\texttt{multiverse\_summary.json} was confirmed on 2026-05-17.

\paragraph{Pre-specification.}
Measure tiers, headline rule, multiverse grid, and the quantity
admissible as the headline reliability index ($\NLRdelta$, with
$\NLRratio$ excluded) were fixed in machine-readable contracts
(\texttt{contracts/} in the repository) committed to version control
before analysis. These contracts function as a de facto preregistration
through Git timestamping; they are not registered on an external
preregistration registry (e.g., OSF).

\paragraph{Second clean execution.}
On 2026-05-17, a second clean execution of the locked pipeline
(\texttt{python -m pipeline.run\_pipeline --mode final --seed 42})
against SHA-256-verified raw archives produced
\texttt{per\_measure\_results.csv} byte-identical to the authoritative
2026-05-09 archive after newline normalisation (LF vs CRLF; the
\texttt{data/processed/} CSVs are not pinned in
\texttt{.gitattributes}, so Git's autocrlf handling can change line
endings without altering content). The numerical content, including
all 50 estimable primary point estimates, confidence intervals, and
the headline pass count, was bit-identical. The multiverse point
estimates ($1{,}344$ specification$\,\times\,$measure cells) were
likewise reproduced with $100\%$ numerical identity to the
authoritative 2026-05-12 archive.

\paragraph{Reproducibility scope.}
The publicly released Zenodo archive contains the full analysis code,
contracts, and verification scripts, but does not redistribute the
Hedge or Clark raw data (which are publicly available from their
original authors). Cloning the public repository and running
\texttt{python scripts/verify.py --mode smoke} executes 12 of 16
checks (R1--R12); the four raw-data checks (R13 hash verification,
R14 ingest evidence, R15 synthetic-data check, R16 long\_df SHA-256)
require the user to first download the original archives and place
them at \texttt{data/raw/hedge.zip} and \texttt{data/raw/clark.zip}.
With those archives present, \texttt{--mode final} executes all 16
checks and reproduces the authoritative outputs byte-identically
(after newline normalisation, as noted above).

\paragraph{Open practices.}
The analysis code, claim contracts, processed outputs, and execution
provenance are publicly archived on Zenodo and GitHub. The raw datasets
are publicly available from their original authors and are referenced
by SHA-256 hashes in \texttt{expected\_hashes.json}.

\paragraph{Conflicts of interest.}
The author declares no conflicts of interest.

\paragraph{Funding.}
No external funding supported this work.

\section*{Author Contributions (CRediT)}

\textbf{Maria Westrin:}
Conceptualization, Methodology, Software, Validation, Formal analysis,
Investigation, Data curation, Writing -- original draft, Writing --
review \& editing, Visualization, Project administration.

\section*{Acknowledgments}

The author thanks the original authors of the Hedge and Clark datasets
for their commitment to open data, without which secondary
methodological work of this kind would be impossible.

\bibliography{references}

\appendix

\section{Multiverse specification grid}
\label{app:specs}

The full $4\times 2\times 3 = 24$-cell grid is the Cartesian product of:
\begin{itemize}[leftmargin=*]
\item $k \in \{3,4,5,6\}$ (KSG nearest-neighbour parameter);
\item correlation method $\in \{\text{Pearson},\ \text{Spearman}\}$
      for the Gaussian baseline;
\item $n_{\min} \in \{10, 15, 20\}$ (minimum estimable sample size).
\end{itemize}
The default specification is $(k=4,\ \text{Pearson},\ n_{\min}=10)$.

\section{Reproducibility checklist}
\label{app:repro}

\begin{enumerate}[leftmargin=*]
\item Raw archives downloaded from the original authors' repositories.
\item SHA-256 of each archive matched the value recorded in
      \texttt{expected\_hashes.json}.
\item Pipeline executed in final mode with $B = 5000$ bootstrap
      replicates and base seed $42$.
\item Verifier script (\texttt{scripts/verify.py --mode final})
      passed all $16$ checks (R1--R16).
\item Per-measure CSV, multiverse CSV, multiverse JSON summary,
      provenance JSON, and release-gate JSON were copied to
      authoritative snapshots dated $2026$-$05$-$09$.
\item Final UTF-8 verification log was scrubbed of local file paths.
\end{enumerate}

\end{document}